# Novel Technique for Volatile Optical Memory Using Solitons

**Mohd Abubakr**
*Student member,
Electronics and Communication Engineering,
Gokaraju Rangaraju Inst. Of Engg & Tech.
Hyderabad, India.*
Email: mohdabubakr@gmail.com

**R. M. Vinay**
*Student member,
Electronics and Communication Engineering,
Gokaraju Rangaraju Inst. Of Engg & Tech.
Hyderabad, India.*
Email: manikyavinay@yahoo.com

**ABSTRACT**
*Recent advances in optics have shown that solitons have a great potential for upgrading the future optical systems which demand fast and reliable data transfer. Along side Different architectures have evolved to realize an optical computer. Due to difference in processing speeds of optical and electronic devices and usage of opto-electronic converters the efficiency of the optical computers is affected. In this paper we bring a novel technique for volatile optical memory using solitons which would be part of realizing an optical computer. Also the principle of read and write operations in the soliton based volatile memory systems is shown.*

## 1.0 INTRODUCTION

The communication revolution of this era demands fast and reliable data transfer from transmitter to receiver. Optical networks have become an essential requirement in keeping the pace of development in terms of transmission of data, voice and multimedia. Already the industry has responded to this change and various technologies have emerged in optics. The undersea fiber optic connections to ultra-long haul connections indicate the impact of this revolution. Photonics promises to revolutionize the area of information technology and the 21st century, just as electronics revolutionized the past hundred years or so [1-2].

Almost all of voice and data (internet) traffic is routed through terrestrial and submarine optical fiber links, connecting the world together. Invention of the optical amplifiers (OAs) and wavelength-division multiplexing (WDM) technology enabled very high capacity optical fiber communication links that run for thousands of kilometers without any electronic repeaters, but at the same time brought many design challenges. In WDM optical fiber communications, information bits are used to modulate the (light) carriers at many wavelengths, which are then transmitted in a single strand of fiber [3].

One of the emerging technologies in optical networks is solitons which promises to benefit the commercial ultra-long haul all-optical multi-terabit networks spanning distances up to many millions of kilometers. Solitons are localized nonlinear waves that have highly stable properties that allow them to propagate very long distances with very little change [4-14].

In this paper we present a novel technique of memory storage using soliton waves. This technique reduces the expenditure spend in optical networks for memory storage. Also the memory techniques we present is volatile in nature, hence could be used in realization of all optical networks without the usage of electronic memory storage systems.

## 2.0 SOLITON WAVES

A soliton was first observed by Russell as a water surface wave in 1844 Korteweg and de Vries in 1895 derived known as the KdV equation, a model equation describing a far-field property of the surface wave in the lowest order of dispersion and nonlinearity. When the equation was numerically solved in a periodic boundary condition by Zabusky and Kruskal in 1965 a set of solitary waves was found to emerge and stably pass each other. They named these solitary waves solitons because of their stability. The term "soliton" was justified when the KdV equation was solved analytically by means of inverse scattering transform (IST) and the solution was described by a set of solitons. Solitons are regarded as a fundamental unit of mode in nonlinear dispersive medium and play a role similar to the Fourier mode in a linear medium. In particular, a soliton being identified as an Eigen value in IST supports its particle (Fermion) concept.

Meanwhile, self-focusing in a Kerr medium was demonstrated and a spatially localized solution analogous to a soliton was found to emerge by the balance of the cubic nonlinearity and refraction. The model equation, called the nonlinear Schrödinger equation, was later found to be integrable by Zakharov and Shabat, also by means of the inverse scattering transform, and the solution is given by a set of solitons and dispersive waves. The theory warrants the stability of the nonlinear Schrödinger solitons.

**3.0 SOLITONS FEATURES**

The monolithic integration of optoelectronic integrated circuits in making rapid progress but due to technical hurdles, are becoming incompatible in optical networks. The technical hurdles include the switching timing of transistor, which is of minimum 5ps, and the processing speed of 5ns. Also the cost of these IC's plays a negative role in building a optical network whereas the theoretical speed of optical fibers is 25THz. At the same time we witnessed the development of new technology called as solitons which had the ability to resolve the problems faced by optical networks.

In 1973 A. Hasegawa and F. D. Tappert, predicted the transmission of light through an optical fiber is similar to the flow of soliton waves [4]. Any optical pulse, which is transmitted into a loss less fiber, forms itself to become solitons like Fourier transmission modes in a linear transmission system. This result was confirmed experimentally in 1980 by [5]. In 1988 Mollenauer and Smith were successful in demonstrating the first all-optical soliton transmission by use of the Raman process in fiber over a distance of 4000 km. They used a 55-ps soliton pulse that was periodically amplified by Raman pumps injected into the fiber every 41.7 km in both the parallel and anti-parallel directions [15-16].

Due to the short pulse duration and high stability, solitons could form the high-speed communication backbone of tomorrow's information super-highway. One of the key technological developments that make use of such soliton pulses for our future cost-effective and repeater less communication is the invention of erbium doped fiber amplifier (EDFA)[2].

The optical transmission losses such as dispersion and non-linearity can be eliminated by transmitting Ultra short pulses. The decrease in the amplitude of the wave can also be eliminated by using Raman amplification or EDFA along with self induced transparency.

In particular, the use of 40 Gbit/s line is becoming the standard for next generation systems and dispersion managed solitons are now believed to be the major candidate. In addition to their application to transmission between two points, solitons have particular merit for use in high speed optical time division networks because of their intrinsic short pulse structure and the stationary pulse shape. Such applications are now being pursued in various parts of the world. Thus, most internet traffics in the 21st century will be carried by optical solitons [17].

Since solitons do not suffer distortion from nonlinearity and dispersion, which are inherent in fibers, the natural next step is to construct an all-optical transmission system in which fiber loss is compensated by amplification. In fact, the optical transmission systems prior to 1991 required repeaters periodically installed in transmission lines to regenerate optical pulses which have been distorted by fiber dispersion and loss. A repeater consists of a light detector and light pulse generators. Consequently, it is the most expensive unit in a transmission system and also the bottleneck to increase in the transmission speed.

Repeater less transmission is a very innovative concept that few people believed to be possible. In the absence of realistic optical amplifiers, Hasegawa in 1983 proposed to use the Raman gain of the fiber itself. The idea was used in the first long distance all-optical transmission experiment by Mollenauer and Smith in 1988. However, Mollenauer succeeded in the 4000 km repeater less transmission experiment. This achievement also leads to the concept of contemporary repeater less optical transmission systems, with and without solitons. This experimental result has attracted serious interest in the optical communication community. In particular, the invention of erbium doped fiber amplifiers (EDFA) have enhanced the concept of all-optical transmission systems to more realistic levels as initiated by Nakazawa et al. in the first reshaping experiment of solitons.

**4.0 REALISATION OF AN OPTICAL COMPUTER**

Realizing an optical computer has been a long time goal in optical industry. Different architectures for optical processors have already been established and implemented at various labs. Various optical processors like labyrinth, SWAS, DOC, AOS, etc have successfully demonstrated the strength and speed of optical processing. The optical processor speed was reduced due to the time taken by fetching the instructions and data from the conventional main memory. The conventional main memory uses the electronic design and opto-electronic converters are used to send and receive data from main memory and processor. The efficiency of the optical system can be further increased if the electronic devices are replaced by optical devices which minimize or avoid the use of opto-electronic converters.

One of the electronic devices is RAM which is very essential in determining the speed of the processor. We bring forward a novel technique for RAM implementation using the soliton concept. This technique is very effective and doesn't require any opto-electronic converter.

**5.0 VOLATILE OPTICAL MEMORY**

We present here a novel technique for Random access memory (RAM) for realization for an optical computer. RAM plays a major role in determining the speed of the computer. To match up to the speed of optical processor the data fetch timings for a RAM should be fast. This condition demands for the RAM which uses the concept of photonics rather than the electronics. We use the established equations of propagation of soliton waves to build a volatile memory system for an optical processor.

Consider a flat sheet of fiber optic material with variable refractive index such that whenever a soliton pulse enters into it, it direction of propagation forms a closed loop. Generally in graded index fibers the direction of the pulse continuously changes with the change in the refractive index of the fiber similarly in this flat sheet of optic material the refractive index changes such that the optical pulse makes a closed loop.

The pulse gets trapped in the closed loop due to the internal reflection. Considering the stable properties of the solitons, the optical pulse used in this process is solitons. Such a trapped soliton pulse can be interpreted as a smallest unit of the optical memory (opbit). In a simple system presence of the soliton can be interpreted as logical '1' and the absence of it as '0'.

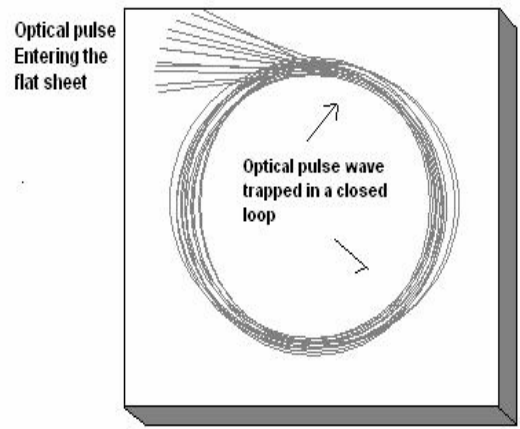

Fig 1: An optical pulse trapped in an optical material.

With the further advance in the technology, each opbit can represent more than two states. This can be realized by taking into account different parameters such as amplitude, phase etc into account.

We confine the analysis only up to two states of the opbit. Taking the standard frequency of the soliton pulse for storing the information as 10THz and assuming that the circumference of the loop is equal to the wavelength of the pulse, we obtain following results. The optical material is manufactured such that the loop formed by the trapped solitons forms a circle or any standard closed curve.

For a 10THz wave, the circumference of the loop is approximately 30μm and the area of the loop is $7.15 \times 10^{-11}$ m$^2$. For practical reasons, the area of the loop is taken as $10^{-10}$ m$^2$ which also accounts for the isolation gap between two adjacent opbits. Therefore a 1cm$^2$ of the flat optical material contains an optical memory storage capacity of more than $10^6$ opbits. Similarly the optic material can contain several layers of such memory storage. A 1 cm$^3$ of material can contain a capacity of approximately $10^{12}$ opbits. Such large dynamic RAM will be compatible with the processing speed of the optical processor.

The read and write operations can be done through a standard laser diode of 10THz frequency (in this case). For read operations the data can be detected using the Boolean AND technique shown in [2]. A standard soliton pulse is sent to the location where the data is present and 'Boolean AND' operation is performed with the data. If the output is '1' then data is equal to '1' and vice-versa.

### 6.0 CONCLUSIONS

Though solitons are in the emerging state, the capabilities of solitons are enormous compared to any other technology. We presented in this paper a novel technique for the volatile memory which is immensely helpful in making a fast optical computer. With a proper engineering design of the optical materials this type of soliton memory can easily be built.

### 7.0 ACKNOWLEDGEMENTS

The authors would like to thank Prof. A.P.N. Rao, Head of Electronics and Communication Dept., GRIET for his valuable suggestions on soliton models.